\documentclass[a4paper]{article}

\usepackage{INTERSPEECH2022}

\title{An Evaluation of Three-Stage Voice Conversion Framework for Noisy and Reverberant Conditions}
\name{Yeonjong Choi, Chao Xie, Tomoki Toda}
\address{
  Nagoya University, Japan
 }
\email{\{yeonjong.choi,xie.chao\}@g.sp.m.is.nagoya-u.ac.jp, tomoki@icts.nagoya-u.ac.jp}

\begin{document}

\maketitle
\begin{abstract}
This paper presents a new voice conversion (VC) framework capable of dealing with both additive noise and reverberation, and its performance evaluation. There have been studied some VC researches focusing on real-world circumstances where speech data are interfered with background noise and reverberation. To deal with more practical conditions where no clean target dataset is available, one possible approach is zero-shot VC, but its performance tends to degrade compared with VC using sufficient amount of target speech data. To leverage large amount of noisy-reverberant target speech data, we propose a three-stage VC framework based on denoising process using a pretrained denoising model, dereverberation process using a dereverberation model, and VC process using a nonparallel VC model based on a variational autoencoder. The experimental results show that 1) noise and reverberation additively cause significant VC performance degradation, 2) the proposed method alleviates the adverse effects caused by both noise and reverberation, and significantly outperforms the baseline directly trained on the noisy-reverberant speech data, and 3) the potential degradation introduced by the denoising and dereverberation still causes noticeable adverse effects on VC performance.

\end{abstract}

\section{Introduction}

Voice conversion (VC) \cite{Sisman21}\cite{Mohammadi17} is a study of converting one’s voice (a source speaker's voice) to sound like another voice (a target speaker's voice) while retaining the linguistic contents. This technique can be applied to many real-life applications, such as personalized speech synthesis, communication aids for the speech-impaired, and voice dubbing for movies \cite{Sisman21}\cite{Mohammadi17}. Thanks to the recent advancements in deep learning methods, VC embraces significant improvements in terms of naturalness and speaker-similarity \cite{Yi20}.

On the other hand, current VC studies mostly require clean speech data in particular from the target speaker, which is hardly guaranteed in real-world scenarios. Additive noise or reverberation can easily interfere speech data available in practical. However, only a small part of the works have been dedicated to VC in noisy and reverberant conditions. For example,  Takashima \textit{et al.} \cite{Takashima13} proposed a sparse representation-based VC using nonnegative matrix factorization to filter out the noise. Miao \textit{et al.} \cite{Miao20} proposed a noise-robust VC that introduces two filtering methods at the pre- and post-processing stages, respectively, to suppress the noise. Despite the contributions of these previous studies, there is a limitation that those methods still need to use clean target speech data during the training stage. It is not always possible to get a large amount of clean target data since it costs a lot to collect them, and sometimes we are only available to use noisy-reverberant speech data. One possible approach to deal with this issue is zero-shot VC \cite{Mottini21}\cite{Huang21}, but its performance tends to degrade compared with VC using sufficient amount of target speech data. Since sufficiently large amount of noisy-reverberant speech data will be available in real scenarios, it is worthwhile to develop a VC framework capable of directly using it.

To deal with this issue, our goal is to establish a noise- and reverberation-robust VC system, especially, where clean target data are unavailable. Recently, Xie \textit{et al.} \cite{Xie21}\cite{Xie21-2} have developed a noisy-to-noisy (N2N) VC framework, where the conversion is achieved while retaining the background additive noise. They utilize a pre-trained speech enhancement (SE) model for separating noisy speech into speech part and noise part. After processing all the noisy training data, they train vector-quantized variational autoencoder (VQ-VAE) VC model \cite{Niekerk20}, in either the VC model will generate clean converted speech, or generate noisy converted speech by adding the separated noise later.

In this paper, we propose to add a dereverberation model to the previous work \cite{Xie21}, so that the frameworks can deal not only with noise, but also with reverberation. Since we assume that the clean VC training data are not available, we first apply denoising and dereverberation process to noisy-reverberant speech data, and then, we train the VQ-VAE-based VC model using the processed data. For a dereverberation model, we utilize the models which are pretrained beforehand with the publicly available mega dataset. Objective and subjective evaluations are conducted for our proposed method. The experimental results show that using both the denoising- and dereverberation-models can process noisy-reverberant speech with acceptable quality, and the VC model training using the denoised and dereverbered speech data is effective for alleviating the adverse effects caused by using noisy-reverberant speech. Our contributions of this work are as follows :
\begin{itemize}
\item We propose a three-stage VC framework consisting of denoising, dereverberation, and VC, using the pre-trained denoising and dereverberation models and non-parallel VC model based on VQ-VAE, making it possible to use noisy-reverberant speech data without the need for clean speech data.
\item We investigate the adverse effects caused by using noisy-reverberant speech data on VC performance.
\item We evaluate how a combination of the denoising and dereverberation preprocessings dedicate to the VC downstream.
\end{itemize}

\begin{figure}[t]
  \centering
  \includegraphics[width=\linewidth]{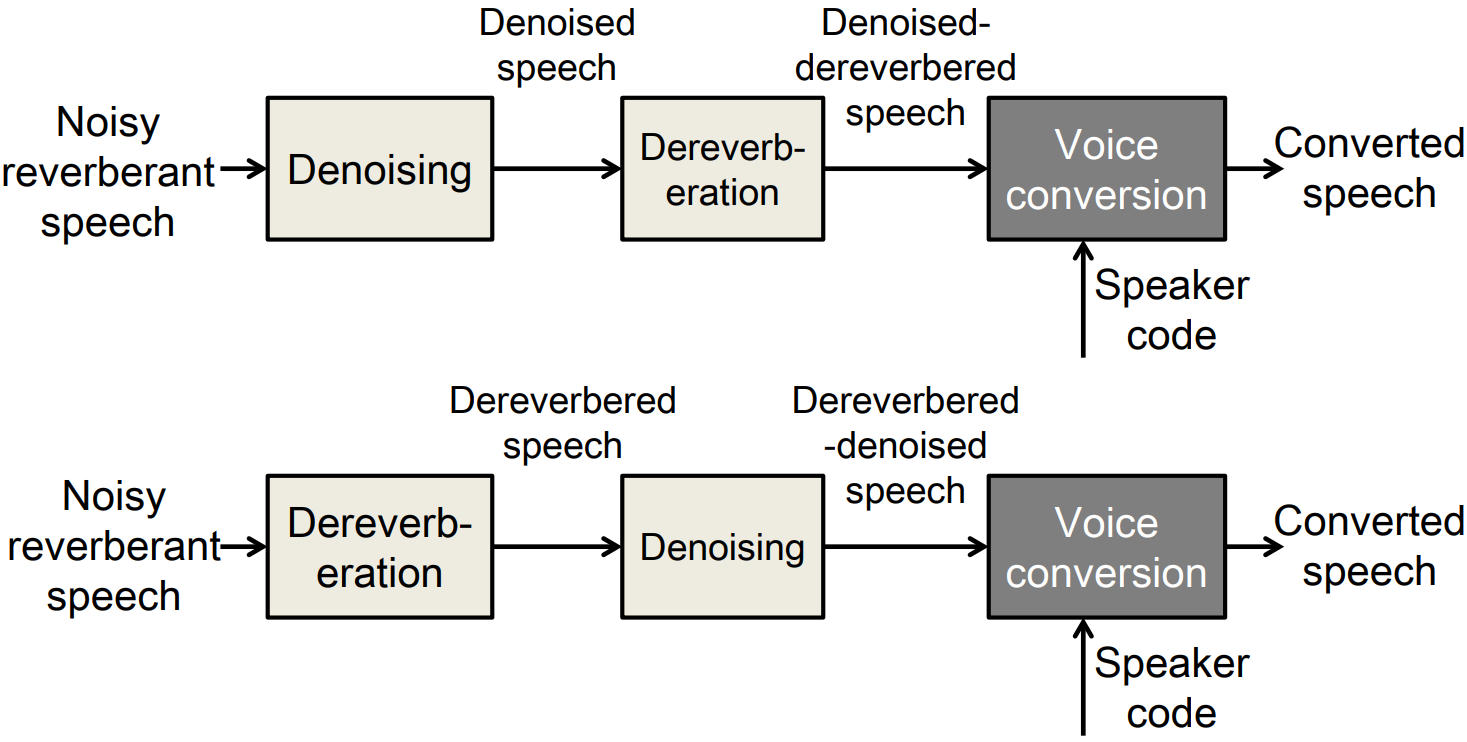}
  \vspace*{-3mm}
  \caption{Overview of the proposed three-stage VC framework for noisy and reverberant conditions. Pretrained denoising- and dereverberation-models are used. Either denoising or dereverberation can be done first, and then the other later.}
  \label{fig:overview}
    \vspace*{-3mm}
\end{figure}

\section{Speech enhancement processing}

Following Xie \textit{et al.} \cite{Xie21}, we use deep complex convolution recurrent network (DCCRN) \cite{Hu20} for the denoising model. It has shown the state-of-the-art performance at the real-time track in Deep Noise Suppression Challenge 2020 \cite{Reddy20}. For dereverberation models, as well as DCCRN, we introduce Time-domain Audio Separation Network (TasNet) \cite{Luo18-1}\cite{Luo18-2}, which have shown in the literature that it outperforms deep long short-term memory (LSTM) baseline that works on time-frequency domain signal. We also introduce Weighted Prediction Error (WPE) dereverberation algorithm \cite{Nakatani10} as one of the typical dereverberation methods based on a signal processing approach instead of a deep learning approach.

\subsection{WPE-based dereverberation}

WPE \cite{Nakatani10} is a signal processing based algorithm that blindly dereverberates acoustic signals based on the long-term linear prediction. The reverberation tail of the input reverberant signal is estimated, and subtracted from the input signal to obtain the optimal estimate of the direct path signal in a maximum likelihood sense \cite{Drude18}. While this algorithm can be applied to multichannel applications, we use it as a single-channel dereverberation method in this paper.

\subsection{Deep denoising- and dereverberation-models}

\subsubsection{DCCRN}

DCCRN \cite{Hu20} is a convolutional recurrent network (CRN) based model designed for single-channel speech enhancement. The encoder and decoder consist of stacked two-dimensional convolution (Conv2D) blocks. Each Conv2D block consists of a convolution layer for the encoder, and a deconvolution layer for the decoder, followed by batch normalization and activation function. LSTM layers are implemented between encoder and decoder for modeling the temporal dependencies. The DCCRN has been shown the state-of-the-art performance on speech enhancement task, by handling the problems of complex calculation. Specifically, complex convolution neural network, complex batch normalization layer, and complex LSTM are introduced, so that the DCCRN can model the correlation between magnitude and phase. In our work, we use a scale-dependent signal-to-distortion (SD-SDR) loss \cite{Roux19} as an objective function for training.

\subsubsection{TasNet}

TasNet \cite{Luo18-1}\cite{Luo18-2} is a neural network that directly models the time-domain waveform signal using a convolutional encoder-decoder architecture, and performs the speech enhancement on the output of the encoder. TasNet consists of an encoder that has a non-negativity constraint on its output, and a linear decoder that converts the encoder output into the waveform signal. Deep LSTM enhancement network performs the enhancement step by estimating a proper weighting function for the encoder output at each time step. TasNet's performance and its effectiveness have been demonstrated by being compared with a number of systems that work on time-frequency domain representations. Like DCCRN, we use the SD-SDR loss as an objective function for training.

\section{Three-stage VC framework for noisy and reverberant conditions}

\subsection{Framework overview}

Let \(s(t)\), \(n(t)\), and \(h(t)\) denote clean speech, additive backgroud noise, and room impulse response (RIR), respectively. The noisy-reverberant speech \(y(t)\) can be noted as
\begin{equation}
y(t) = s(t) * h(t) + n(t)
\end{equation}
where * stands for the convolution operator. Using \(y(t)\) directly for VC model training- and conversion-stage would be result in degraded synthesized
speech, leading to the poor VC performance. Instead, it would be ideal if we could estimate \(s(t)\) from \(y(t)\), by removing \(n(t)\) and \(h(t)\) separately \cite{Maciejewski20}.

In our framework, as illustrated in Figure \ref{fig:overview}, the noisy-reverberant speech data are preprocessed by a pretrained denoising- and a dereverberation-model, respectively, before feeding to the VC downstream. Two SE models are pretrained on the publicly available mega dataset \cite{Reddy20}\cite{Maciejewski20}, and fixed during the VC training. In conversion stage, the same preprocessing models can be used for noisy-reverberant input data. In this paper, we also investigate how the order of the preprocessings (whether denoising is first, or dereverberation is first) will affect the SE and VC downstream performances.

\subsection{Voice conversion model}

Following Xie \textit{et al.} \cite{Xie21}, we use VQ-VAE for the VC model \cite{Niekerk20}. VQ-VAE consists of an encoder, a bottleneck layer, and a decoder. The encoder consists of five convolution blocks composed of a one-dimensional convolution (Conv1D) layer followed by a batch normalization layer and an activation function. Given log mel-spectrogram sequence \(\{\mathbf{x}_t | t = 1, ..., T\}\) as input, the encoder generates a sequence of latent vectors \(\{\mathbf{z}_j | j = 1, ..., N\}\), and this latent vectors are sent to a vector-quantized (VQ) bottleneck, which consists of a trainable codebook \(\{\mathbf{e}_i | i = 1, ..., 512\}\), where \(\mathbf{e}_i\) is a 64-dimensional vector. Each of the latent vectors of the encoder \(\mathbf{z}_j\) is mapped into the nearest vector \(\mathbf{e}_k\) of the codebook by :

\begin{equation}
\mathbf{\hat{z}}_j = \mathbf{e}_k
\end{equation}
\begin{equation}
k = \arg\min_{i}||\mathbf{z}_j - \mathbf{e}_i||^2
\end{equation}
where \(\{\mathbf{\hat{z}}_j | j = 1, ..., N\}\) is a replaced discrete latent representation. The decoder consists of a lightweight recurrent network to reconstruct the waveform based on \(\mathbf{\hat{z}}_j\) and the embedded speaker identity (speaker code) in an autoregressive way.

During the training stage, the VQ-VAE is trained to minimize the sum of the negative log-likelihood of the reconstruction loss and the commitment loss :

\begin{equation}
\mathcal{L} = -\frac{1}{T}\sum_{t=1}^{T}\log p(\mathbf{x}_t|\mathbf{\hat{z}}) + \beta \frac{1}{N}\sum_{j=1}^{N}||\mathbf{z}_j - \text{sg}(\mathbf{\hat{z}}_j)||^2
\end{equation}
where \(\beta\) is the commitment weight, and \(\text{sg}(\text{·})\) indicates the stop-gradient operation. Since this function is not differentiable, the gradient of the loss function through the codebook is approximated by the straight-through estimator \cite{Benigo13}. To encourage the bottleneck layer to disentangle speaker information and linguistic information during training, we condition the decoder on a speaker code. The denoised and dereverberant speech dataset is used as a nonparallel training dataset.

\section{Experimental Evaluations}

\subsection{Dataset}

All the experiments were conducted on single-channel, 8kHz sampled speech data.

For the training of the denoising model, we used DNS Challenge 2020 dataset \cite{Reddy20}. It contained 500 hours of speeches from 2,150 speakers, and 65,000 noise clips in 150 classes. 6,000 speech clips and 500 noise clips were randomly sampled as the validation data. The SNR levels were set between 5 and 20 dB.

For the training of the dereverberation models, we used WHAMR! dataset \cite{Maciejewski20}, which was originally designed for speech separation under noisy-reverberant conditions. It contained 58.03 hours of training speech, and 14.65 hours for the validation set. For the dereverberation models training, we used \texttt{s1\_reverb} set as the input signals, which is a set of reverberated single speaker sources, and used \texttt{s1\_anechoic} set as the ground truth signals, which is a clean set. Since the clips consist of 2 channels originally, we extracted the left channel of those to make it single channel sources.

We used VCC2018 dataset \cite{Lorenzo-Trueba18} for VQ-VAE VC models training and evaluation. There were 8 source speakers and 4 target speakers in the data, where each speaker uttered 81 training clips and 35 evaluation clips. For noise data, PNL100 nonspeech sounds \cite{Hu10} were used, which consist of 100 clips of environmental records. We sampled the noise clips from N1 to N85 to mix with VCC2018 training set at eight SNR levels (6, 8, 10, 12, 14, 16, 18, 20 dB), and N86 to N100 for evaluation set at four SNR levels (7, 11, 15, 19 dB). For our experimental purpose, we generated noisy and reverberant versions (NR-VCC2018) as follows. We first generated reverberant versions of VCC2018 and PNL100 by convolving with room impulse responses (RIR). Here, we followed the RIR settings of evaluation set of WHAMR! \cite{Maciejewski20}, so those RIRs were not seen by dereverberation model during its training. We divided the RIR settings into two groups, one to be convolved with training sets of VCC2018 and PNL100, and the other with evaluation sets. Note that different RIRs were used for generating reverberant speech and reverberant noise. After generating reverberant VCC2018 and PNL100, those were mixed to generate the final NR-VCC2018. For evaluation data, four speakers (VCC2SM3, VCC2SM4, VCC2SF3, VCC2SF4) were selected as source speakers, and two speakers (VCC2TF2, VCC2TM2) as target speakers.

\subsection{Model training}

\subsubsection{Speech enhancement models}

For denoising-DCCRN, we used the same settings as \cite{Xie21}. For dereverberation models (TasNet, DCCRN), we used Asteroid toolkit \cite{Pariente20}. We basically followed the default settings of \texttt{egs/whamr/TasNet} and \texttt{egs/librimix/DCCRNet} for TasNet and DCCRN, respectively. For both of TasNet and DCCRN, Adam was used as the optimizer, and the early stopping mechanism was introduced to choose each optimized model. For WPE, we used NARA-WPE package \cite{Drude18}.

\subsubsection{VC models}

We used a VQ-VAE VC model that was implemented by \cite{Niekerk20}. The window length, hop size, and FFT length were set to 20 ms, 5 ms, 1024, respectively. The batch size was 64. The value of \(\beta\) was set to 0.25. Adam was used as the optimizer with an initial learning rate of 2 · 10\textsuperscript{-4}. The learning rate was halved after 300k steps of training. The total training steps were 700k steps.

\subsection{Methods to be evaluated}

In the experiments, our goals are to find out which combination of the denoising- and dereverberation-model will show the best performance for SE, and how it is correlated with the performance of VC downstream. For the experimental comparison, we prepared the following VQ-VAE VC frameworks :
\begin{itemize}
\item NR-dn-dr-VC (trained on first-denoised, later-dereverbered NR-VCC2018)
\item NR-dr-dn-VC (trained on first-dereverbered, later-denoised NR-VCC2018)
\item NR-dn-VC (trained on denoised-only NR-VCC2018, for ablation study)
\item NR-dr-VC (trained on dereverbered-only NR-VCC2018, for ablation study)
\item C-VC (trained on clean VCC2018, as an upper-bound)
\item NR-VC (trained on NR-VCC2018 without any preprocessing, as a lower-bound)
\end{itemize}
Here, "NR" indicates noisy-reverberant data, "C" means clean data, "dn" denotes the denoising model, and "dr" denotes the dereverberation model. Note that there are 3 kinds of variations for dereverberation model (WPE, TasNet, DCCRN), so the "dr" will be denoted as one of "W", "T", and "D", in the result tables.

\subsection{Experimental results}

\subsubsection{Results of objective evaluation on speech enhancement}

We first present the preprocessing performances of the denoising-, dereverberation-model, and their combinations, on NR-VCC2018 training set. The results are shown in Table \ref{tab:dndr}. We can find that if we do either denoising or dereverberation, we can get enhanced results, comapred with input. For dereverberation, in terms of SI-SDR \cite{Roux19}, PESQ \cite{Rix01} , and STOI \cite{Taal10}, deep learning based methods (TasNet, DCCRN) outperform WPE. However, WPE outperforms those in terms of SAR \cite{Vincent06}, since WPE method dereverb the signals by linearly filtering the inputs, generating less artifacts. Among the two deep learning based methods, TasNet performs better in terms of PESQ with the score of 2.00. The combinations of two preprocessings show better performance than using either of denoising- or dereverberation-model, which is expectable. We expected doing denoising first would be better than doing dereverberation first \cite{Maciejewski20}, but from the table, it seems it depends on which model we choose for dereverberation. Among the six combinations, NR-dn-T shows the best results in terms of PESQ with the score of 2.74.

\begin{table}[t]
  \caption{The objective results of preprocessing models on NR-VCC2018 training set.}
  \label{tab:dndr}
  \centering
  \vspace*{-4mm}
  \begin{tabular}{c|c c c c}
    \toprule
    & SI-SDR(dB) & PESQ & STOI & SAR(dB) \\
    \midrule
    NR                & 2.64 & 1.74 & 0.85 & 9.63             \\
    \midrule
    NR-dn                & 3.92 & 2.31 & 0.88 & 14.52            \\
    \midrule
    NR-W            & 3.11 & 1.80 & 0.85 & \textbf{11.54}            \\
    NR-T            & 5.04 & \textbf{2.00} & \textbf{0.88} & 8.76             \\
    NR-D            & \textbf{5.16} & 1.89 & \textbf{0.88} & 8.92             \\
    \midrule
    NR-W-dn       & 4.46 & 2.45 & 0.89 & \textbf{17.04}            \\
    NR-T-dn       & 5.58 & 2.57 & 0.90 & 9.85             \\
    NR-D-dn       & \textbf{6.54} & 2.70 & \textbf{0.91} & 11.50            \\
    NR-dn-W       & 4.19 & 2.36 & 0.88 & 16.30            \\
    NR-dn-T       & 6.00 & \textbf{2.74} & \textbf{0.91} & 10.57             \\
    NR-dn-D       & 6.49 & 2.62 & \textbf{0.91} & 11.55             \\
    \bottomrule
  \end{tabular}
  \vspace*{-3mm}
\end{table}

\subsubsection{Results of objective evaluation on VC}
 
 The mel-cepstral distortion (MCD) \cite{Kubichek93} objective evaluations of C-VC are shown in Table \ref{tab:MCD-C}. Though we assume that we cannot use clean VC training data, we evaluate this model to find out how the errors contained in the preprocessed data will affect the converted speech, when the VC model is trained on clean data. Here, the inputs are evaluation sets that are clean (C), or preprocessed to be clean. For example, when using the denoising model only, inputs to the denoising model are noisy-only data (N-dn). Likewise, when using the dereverberation model only, inputs to the dereverberation model are reverberant-only data (R-dr). For the combinations of two preprocessings, inputs are noisy-reverberant (NR-dn-dr, NR-dr-dn). When doing either denoising or dereverberation, MCD increases compared with when inputs are clean, due to the distortion caused by each preprocessing. When we do both of two preprocessings, MCD increases more, due to the errors accumulated from those models. Dealing with these accumulated errors will be one of our future works.
 
 \begin{table}[t]
  \caption{The objective results of the model C-VC, where input evaluation data are clean, or preprocessed to be clean.}
  \label{tab:MCD-C}
  \centering
    \vspace*{-4mm}
  \begin{tabular}{c|c|c|c c c}
    \toprule
     Input & C & N-dn & R-W & R-T & R-D\\
    \midrule
    MCD(dB) & 7.30 & 7.43 & 7.51 & 7.49 & 7.46\\
    \bottomrule
  \end{tabular}
  \begin{tabular}{c|c c c}
    \toprule
     Input & NR-W-dn & NR-T-dn & NR-D-dn \\
    \midrule
    MCD(dB) & 7.63 & 7.67 & 7.58 \\
    \bottomrule
  \end{tabular}
  \begin{tabular}{c|c c c}
    \toprule
     Input & NR-dn-W & NR-dn-T & NR-dn-D\\
    \midrule
    MCD(dB) & 7.70 & 7.65 & 7.62 \\
    \bottomrule
  \end{tabular}
  \vspace*{-5mm}
 \end{table}

We return to the assumption that clean training data are not available. MCD results of VQ-VAEs trained on various preprocessed data are shown in Table \ref{tab:MCD-P}. Inputs of each model are the evaluation sets that are preprocessed in the same way as the corresponding training set. As in Table \ref{tab:dndr}, doing either of denoising or dereverberation for noisy-reverberant training data can improve VC performance, and this can be further improved by doing all of the preprocessings. We can observe some results are correlated with those in Table \ref{tab:dndr}. For example, for the cases of NR-dr-VC, using TasNet among the three dereverberation models shows the best result, as also shown in Table \ref{tab:dndr}. For the combinations of two prerprocessings, we can find that NR-dn-T-VC is shown to be the best, also following Table \ref{tab:dndr}.

\begin{table}[t]
  \caption{The objective results of VCs trained on various preprocessed data.}
  \label{tab:MCD-P}
  \centering
    \vspace*{-4mm}
  \begin{tabular}{c|c|c|c}
    \toprule
      & MCD(dB) & & MCD(dB)  \\
    \midrule
    NR-VC  & 8.96 &  NR-W-dn-VC & 8.57       \\
    NR-dn-VC    & 8.66 &  NR-T-dn-VC & 8.07       \\
    NR-W-VC     & 9.02 & NR-D-dn-VC & 8.46       \\
    NR-T-VC     & \textbf{8.35} & NR-dn-W-VC & 8.62         \\
    NR-D-VC     & 8.89 & NR-dn-T-VC & \textbf{8.04}        \\
     &  & NR-dn-D-VC & 8.61 \\
    \bottomrule
  \end{tabular}
  \vspace*{-3mm}
\end{table}

\subsubsection{Results of subjective evaluation on VC}

We chose NR-dn-T-VC and NR-T-dn-VC for the subjective evaluations, as they show the best performances in terms of MCD, as well as C-VC and NR-VC as comparison groups. For the naturalness test, we conducted mean opinion score (MOS) test, where 10 listeners, recruited on Amazon Mechanical Turk, were asked to give a naturalness score from 1 to 5 (higher is better). We sampled 6 samples from each of 8 conversion-pairs, converted by 4 systems. We also sampled 8 clean speeches from each 2 target-speakers, thus 208 samples were evaluated in total by every listener. For the speaker-similarity test, we conducted the similarity (SIM) test \cite{Lorenzo-Trueba18}. In the SIM test, each listener listened to two kinds of samples: a converted speech, and a clean speech of the target speaker with different sentence. Listeners were asked to determine whether those samples were uttered by the same speaker: \textit{Definitely the same, Maybe the same, Maybe different, Definitely different}. We sampled 4 samples from each conversion-pairs, thus 128 samples were evaluated in total by every listener.

The results are shown in Table \ref{tab:MOS}. For SIM, the percentages indicate the added percentages of \textit{Definitely the same} and \textit{Maybe the same}. 
We can get much better results by preprocessing the noisy-reverberant VC data, compared with directly using it for the training. But there are still many rooms to improve its performance. Meanwhile, the effect of the order of preprocessing on VC performance seems marginally small.

\begin{table}[t]
  \caption{The MOS and SIM results of the four VC systems with 95\% confidence interval.}
  \label{tab:MOS}
  \centering
    \vspace*{-4mm}
  \begin{tabular}{c|c c}
    \toprule
     & MOS & SIM (\%) \\
    \midrule
      Clean & 4.56 $\pm$ 0.11 & - \\
      \midrule
      C-VC & 3.80 $\pm$ 0.07 & 71.52 $\pm$ 5.04 \\
      NR-dn-T-VC & \textbf{3.07 $\pm$ 0.08} & 47.44 $\pm$ 5.55 \\
      NR-T-dn-VC & 2.95 $\pm$ 0.08 & \textbf{50.32 $\pm$ 5.56} \\
      NR-VC & 1.41 $\pm$ 0.06 & 21.61 $\pm$ 4.59 \\
    \bottomrule
  \end{tabular}
  \vspace*{-5mm}
\end{table}

\section{Conclusions}

In this paper, we presented a new three-stage VC framework based on SE processing using a pretrained denoising model, a dereverberation model, and VC process using a nonparallel VC model based on a VAE. The experimental results showed that noise and reverberation additively cause significant VC performance degradation, the proposed method alleviates the adverse effects caused by noise and reverberation, 
and the potential degradation introduced by the denoising and dereverberation still causes noticeable adverse effects on VC performance. As future works, we plan to test our framework on 16kHz sampled data, compare with other VC frameworks \cite{Mottini21}\cite{Huang21}, and improve VC performance by reducing the potential degradation.

{\bf Acknowledgements:} This work was partly supported by JST CREST Grant Number JPMJCR19A3, Japan.

\bibliographystyle{IEEEtran}


\end{document}